\documentclass[11pt]{article}

\usepackage{graphicx,amsmath,amsfonts,amssymb,dcolumn,amsthm}
\usepackage{slashed}
\usepackage{bbm}                
\usepackage{xspace}
\usepackage{pgffor}	
\usepackage{tikz}
\usepackage{amssymb}
\usepackage{amsmath}
\usepackage{amsfonts}
\usepackage{mathbbol}
\usepackage{mathrsfs}
\usepackage{amssymb}
\usepackage{latexsym}

\numberwithin{equation}{section} 
\begin{document}

\title{\textbf{Gauge field spectrum in massive Yang-Mills theory with Lorentz violation}}
\author{ \textbf{T. R.~S.~Santos$^{1}$}\thanks{tiagoribeiro@if.uff.br}\ , \textbf{R. F.~Sobreiro$^{1}$}\thanks{sobreiro@if.uff.br}\ , \textbf{{A.~A.~Tomaz$^{1,2}$}\thanks{tomaz@cbpf.br}}\\\\
$^1$\textit{{\small UFF $-$ Universidade Federal Fluminense,}}\\
\textit{{\small Instituto de F\'{\i}sica, Campus da Praia Vermelha,}}\\
\textit{{\small Avenida General Milton Tavares de Souza s/n, 24210-346,}}\\
\textit{{\small Niter\'oi, RJ, Brasil.}}\\\\
$^2$\textit{{\small CBPF $-$ Centro Brasileiro de Pesquisas F\'isicas,}}\\
\textit{{\small Rua Dr. Xavier Sigaud, 150 , Urca, 22290-180}}\\
\textit{{\small Rio de Janeiro, RJ, Brasil}}}
\date{}
\maketitle

\begin{abstract}
The spectrum of the massive CPT-odd Yang-Mills propagator with Lorentz violation is performed at tree-level. The modification is due to mass terms generated by the exigence of multiplicative renormalizability of Yang-Mills theory with Lorentz violation. The causality analysis is performed from group and front velocities for both, spacelike and timelike background tensors. It is shown that, by demanding causality, it is always possible to define a physical sector for the gauge propagator. Hence, it is expected that the model is also unitary if one takes the Faddeev-Popov ghost into account.
\end{abstract}

\section{Introduction}\label{intro}

The Abelian sector of the minimal Standard Model Extension (SME) \cite{Colladay:1996iz,Colladay:1998fq, Kostelecky:2003fs}, \textit{i.e.}, Lorentz- and CPT-violating QED, has received considerable attention in the last decades. This model is characterized by some couplings among constant background tensors and gauge and matter fields operators with dimension bounded by four (power-counting renormalizability). At both sectors of this extension, namely, gauge and fermionic sectors, there exist the presence of CPT-even and CPT-odd background tensors. For instance, the CPT-odd gauge sector of the SME \cite{Carroll:1989vb} has shown to be consistent under quantum aspects such as stability, causality, anomaly-freeness, unitarity \cite{Adam:2001ma,Adam:2001kx,Santos:2015koa,Santos:2016bqc}. Furthermore, this sector, when analyzed in the presence of Higgs mechanism, is also causal and unitary for a spacelike background tensor \cite{BaetaScarpelli:2003yd}. Moreover, the matter-free sector of the SME is stable and causal in the low-energy regime, when compared with the Planck scale. However, stability and causality can be spoiled close to the Planck scale \cite{Kostelecky:2000mm}. In fact, since the SME is an effective model that describes low-energy effects of an underlying quantum gravity theory at Planck scale, it is expected that its parameters be Planck suppressed. Nevertheless, the stability and causality violation close to the Planck scale can be avoided with a mechanism of spontaneous Lorentz violation in specific scenarios \cite{Kostelecky:1988zi}. Despite of few complications due to the inversion problem of wave operators, some important results have been obtained to establish the quantum consistency of models with CPT-even Lorentz violating photon sector \cite{Kostelecky:2002hh,Casana:2010nd, Casana:2009xs}. The analysis of the consistency of the CPT-odd photon sector was performed in \cite{Schreck:2011ai}. Also, there are possibilities for mass generation from massive Lorentz violating coefficients of CPT-even, and the study of the consistency of these models was performed in \cite{Gabadadze:2004iv,Fargnoli:2014lca}.

It is worth mentioning that other important quantum aspect of the Lorentz-violating QED, \textit{i.e.}, renormalizability, has been verified at one-loop order \cite{Kostelecky:2001jc}. The extension of the proof to all orders in perturbation theory was performed in \cite{Santos:2015koa,DelCima:2012gb} from algebraic renormalization approach \cite{Piguet:1995er}. An interesting result pointed out in \cite{Santos:2015koa} is the non radiative generation of mass terms for the photon from Lorentz violating coefficients. In fact, the radiative generation of mass terms for the photon would imply on the breaking of the gauge invariance of the theory, namely, the photon propagator would no longer be transverse \cite{Altschul:2003ce}. Moreover, it was formaly shown in \cite{Santos:2015koa} that the Abelian Chern-Simons-like term is not generated by radiative corrections, see also \cite{Jackiw:1999yp,Bonneau:2000ai,DelCima:2009ta} and references therein.

Albeit the non-Abelian sector of the SME has not receiving as many attention as its Abelian version, some studies indicate that this model is also well established at quantum level. In fact, the unitarity and causality of Lorentz-violating $SO(3)$ model is discussed in \cite{Baeta Scarpelli:2006kd}. Moreover, just like the Abelian case, radiative generation of non-Abelian Chern-Simons-like term from CPT-odd of fermionic sector was studied \cite{Gomes:2007rv}. The renormalization at one-loop order of pure Yang-Mills theory with Lorentz violation was performed in \cite{Colladay:2006rk}. Remarkably, the ultraviolet behavior of the CPT-even coupling may provide an upper bound for this coefficient, in contrast to CPT-odd couplings -- this does not happen at the Abelian version, where both, CPT-even and CPT-odd coefficients, have the same behavior at any energy scale. The one-loop renormalization of the Electroweak Sector and QCD with Lorentz violation was performed in \cite{Colladay:2009rb,Colladay:2007aj}, respectively. 

A recent work \cite{Santos:2014lfa} extends the analysis of the renormalizability of pure Yang-Mills theory with Lorentz violation to all orders in perturbation theory. The authors made use of the algebraic renormalization technique along with the external Symanzik sources method \cite{Symanzik:1969ek}. The latter consists in introducing a set of external classical fields in order to control the broken symmetries. The authors used this method due to the following reasons: As we shall see latter, the Lorentz violation is described by a constant background tensor $c_\mu$; a consequence of this is that the gauge symmetry is maintained in a weak form (by neglecting surface terms). Thus, in order to ensure a stronger symmetry, the Becchi-Rouet-Stora-Tyutin (BRST) symmetry \cite{Becchi:1975nq,Tyutin:1975qk} is required by the model. From this symmetry and Symanzik method it is possible to treat the model with Lorentz and BRST symmetries. In this way the applicability of the quantum action principle becomes possible \cite{Lowenstein:1971vf}. Hence, the renormalizability proof of the theory can be established. Moreover, with the BRST formalism, the solutions of the quantum theory are restricted to the solution of a cohomology problem. Remarkably, the Symanzik method together with the BRST formalism induce, in a natural way, vacuum terms and extra mass terms, namely, $M^2A^{a\mu}A^a_{\mu}$ and $V^{\mu\nu}A^a_{\mu}A^a_{\nu}$, where $M$ and $V^{\mu\nu}$ are constants related to $c_\mu$. These massive terms will drastically modify the gauge field propagator, see \cite{Santos:2014lfa}.

We also mention that, since the BRST symmetry of pure Yang-Mills theory with Lorentz violation is well established -- after Symanzik method --, the quantum sector is expected to be free of nonphysical modes \cite{Kugo:1979gm}. However, once the Symanzik sources attain their physical values, the BRST symmetry of the theory is explicitly broken and the theory is driven to a new phase. A direct consequence is that the usual gauge field propagator is modified. It is worth to notice that the BRST symmetry breaking could be interpreted as a spontaneous BRST symmetry breaking \cite{Dudal:2012sb}.

In the present work we focus our analysis on the consistency of the massive pure Yang-Mills theory with CPT-odd Lorentz violation term. In particular we are interested on the spectrum of the gauge field, which is modified by the CPT-odd violating term and massive terms of the type $M^2A^{a\mu}A^a_{\mu}$ and $V^{\mu\nu}A^a_{\mu}A^a_{\nu}$. Hence, by studying the tree-level propagator\footnote{The limitation to the tree-level propagator of our analysis ensures necessary conditions for causality and unitarity. A complete analysis would require the study of microcausality of the theory. Eventually, this would require the computation of complex integrals for the propagators we find, which have a complicated pole structure. This analysis is beyond the scope of the present work.} we analyze the causality (absence of tachyonic states) and partial unitarity (existence of physical states for the gluon). All analysis are performed for both, timelike and spacelike background fields. In essence, the propagator depends on the Lorentz violating scale $\mu$, the momentum scale $k^\mu$, and two dimensionless parameters $\alpha$ and $\beta$, which are related to the massive terms. Hence, we find the relations among these scales and parameters for which the model presents physical gauge states and no tachyonic modes. The absence of tachyonic states is discussed by analyzing the group and front velocities related to the dispersion relations of each state obtained from the propagator. The physical states are analyzed by the saturation of the tree-level propagator and by checking the positive-definiteness of the eigenvalues of the residue matrix in each simple pole of the gauge field propagator. We find that, by demanding causality, there exist physical states. However, ghost modes are always present, which is an expected result since the Faddeev-Popov ghosts are also present at quantum computations.

This work is organized as follows: In Sect.~\ref{YM}, we provide the definitions and conventions of the pure Yang-Mills theory with Lorentz violation. Moreover, the modified renormalizable model with the extra terms -- mass terms -- is provided. In Sect.~\ref{SLC} causality and partial unitarity are analyzed for a spacelike background tensor in different situations for the parameters $\alpha$ and $\beta$. In Sect.~\ref{TLC} the study of the causality and partial unitarity is performed for a timelike background tensor and the parameters $\alpha$ and $\beta$ are treated simultaneously. Our final comments are placed in Sect.~\ref{FINAL}.

\section{Pure Yang-Mills theory with Lorentz violation}\label{YM}

The pure Yang-Mills theory with Lorentz violation, besides the usual Yang-Mills term, includes a sector of Lorentz symmetry breaking\footnote{We will neglect the CPT-even Lorentz breaking sector due to fact that CPT-even terms contribute in a highly nontrivial way to the wave operator, making it virtualy impossible to be inverted.}. This breaking sector is obtained by embedding the three-dimensional Chern-Simons action into four dimensions through the coupling with a background tensor with mass dimension 1. We consider here a theory with $SU(N)$ symmetry group. The gauge fields are algebra-valued $A_{\mu}=A^a_{\mu}T^a$, where $T^a$ are the generators of the $SU(N)$ algebra. They are chosen to be anti-Hermitian and to have vanishing trace. The usual Lie algebra is given by $[T^a,T^b]=f^{abc}T^c$, where $f^{abc}$ are the skew-symmetric structure constants. The Latin indices run as $\left\{a,b,c,\dots\right\}\;\in\;\left\{1,2,\dots,N^2-1\right\}$.

The action that describe the model reads\footnote{We are using the metric tensor as $\eta=diag(+1,-1,-1,-1)$ and the completely skew-symmetric Levi-Civita tensor $\epsilon^{\mu\nu\alpha\beta}$ is normalized as $\epsilon^{0123}=+1$.}
\begin{eqnarray}
\Sigma_0&=&\Sigma_{YM}+\Sigma_{LVO}\;,
\label{1}
\end{eqnarray}
where
\begin{eqnarray}
\Sigma_{YM}&=&-\frac{1}{4}\int d^4x\;F^a_{\mu\nu}F^{\mu\nu a}
\label{2}
\end{eqnarray}
is the classical Yang-Mills action. The field strength is defined as
\begin{eqnarray}
F^a_{\mu\nu}&=&\partial_{\mu}A^a_{\nu}-\partial_{\nu}A^a_{\mu}-gf^{abc}A^b_{\mu}A^c_{\nu}\;.
\label{3}
\end{eqnarray}
The CPT-odd Lorentz violating sector is
\begin{eqnarray}
\Sigma_{LVO}&=&-\frac{1}{2}\int d^4x\;\epsilon^{\mu\nu\alpha\beta}c_{\mu}\left(A^a_{\nu}\partial_{\alpha}A^a_{\beta}+\frac{g}{3}f^{abc}A^a_{\nu}A^b_{\alpha}A^c_{\beta}\right)\;,
\label{4}
\end{eqnarray}
where $c_{\mu}=\mu v_{\mu}$, $\mu$ is a mass parameter and $v_{\mu}$ is a fixed four-vector of unit length which selects a preferred direction in the spacetime. The action \eqref{1} is invariant under exchange of reference frames, \textit{i.e.}, observer Lorentz transformations. Notwithstanding, the Lorentz symmetry is missed under particle Lorentz transformations, where the reference systems keep the same.

As it was shown in \cite{Santos:2014lfa}, the BRST quantization and Symanzik trick \cite{Symanzik:1969ek} allow the introduction of a Proca mass term along with a mixing mass term in the Yang-Mills theory with Lorentz violation that, from algebraic renormalization approach, ensure the renormalizability of the model. The additional action is
\begin{eqnarray}
\Sigma_{M}&=&-\int d^4x\left[\mu^2\left(3\alpha+2\beta\right)v^2 A^a_\mu A^{\mu a}-2\mu^2\beta v^\mu v^\nu A^a_\mu A^a_\nu\right]\:,
\label{5A}
\end{eqnarray}
where $\alpha$ and $\beta$ are dimensionless coefficients and $v^2=v_\mu v^\mu$. Thus, the complete action of the model is 
\begin{eqnarray}
\Sigma&=&\Sigma_0+\Sigma_{M}+\Sigma_{gf}\:,
\label{5}
\end{eqnarray}
where the action $\Sigma_{gf}$ is the gauge fixing action, which is needed to compute the gauge field propagator. The action \eqref{5} has the explicit form
\begin{eqnarray}
\Sigma&=&-\frac{1}{4}\int d^4xF^a_{\mu\nu}F^{\mu\nu a}-\frac{\mu}{2}\int d^4x\epsilon^{\beta\mu\nu\alpha}v_{\beta}\left(A^a_{\mu}\partial_{\nu}A^a_{\alpha}+\frac{g}{3}f^{abc}A^a_{\mu}A^b_{\nu}A^c_{\alpha}\right)+\nonumber\\
&-&\int d^4x\left[\mu^2\left(3\alpha+2\beta\right)v^2 A^a_\mu A^{\mu a}-2\mu^2\beta v^\mu v^\nu A^a_\mu A^a_\nu\right]-\frac{1}{2\xi}\int d^4x(\partial_\mu A^{a\mu})^2\:,
\label{6}
\end{eqnarray}
and $\xi$ is the gauge fixing parameter. Considering only bilinear terms in the gauge field, the  action \eqref{6} in the momentum space has the form
\begin{eqnarray}
\Sigma_{quad}&=&\frac{1}{2}\int \frac{d^4k}{(2\pi)^4}\;A_a^{\mu}(k)\mathcal{O}^{ab}_{\mu\nu}A_b^{\nu}(-k)\;,
\label{8}
\end{eqnarray}
where 
\begin{eqnarray}
\mathcal{O}^{ab}_{\mu\nu}&=&\delta^{ab}\left\{-\left(k^2-\mu^2\Delta v^2\right)\theta_{\mu\nu}-\left(\frac{k^2}{\xi}-\mu^2\Delta v^2\right)\omega_{\mu\nu}+\mu S_{\mu\nu}+\mu^2\Omega\Lambda_{\mu\nu}\right\}\;.
\label{9}
\end{eqnarray}
The operator $\mathcal{O}^{ab}_{\mu\nu}$ is the wave operator while $\Delta=-6\alpha-4\beta$ and $\Omega=4\beta$. The operators $\theta$ and $\omega$ are the usual transversal and longitudinal projectors, namely,
\begin{eqnarray}
\theta_{\mu\nu}&=&\eta_{\mu\nu}-\frac{k_{\mu}k_{\nu}}{k^2}\;,\nonumber\\	
\omega_{\mu\nu}&=&\frac{k_{\mu}k_{\nu}}{k^2}\;,
\label{10}
\end{eqnarray}
while,
\begin{eqnarray}
S_{\mu\nu}&=&i\epsilon_{\mu\nu\alpha\beta}v^{\alpha}k^{\beta}\;,\nonumber\\
\Lambda_{\mu\nu}&=&v_{\mu}v_{\nu}\;.
\label{11}
\end{eqnarray}
In order to compute the inverse of the wave operator -- following \cite{BaetaScarpelli:2003yd} -- we need to define the following extra operator
\begin{equation}
\Sigma_{\mu\nu}=v_{\mu}k_{\nu}\;.\label{11a}
\end{equation}
The operator $\Sigma_{\mu\nu}$, together with the operators shown in \eqref{10} and \eqref{11}, form a closed algebra. The operator algebra is displayed in Table \ref{tableA1}, where
\begin{eqnarray}
f_{\mu\nu}&\equiv&(v^2k^2-\lambda^2)\theta_{\mu\nu}-\lambda^2\omega_{\mu\nu}-k^2\Lambda_{\mu\nu}+\lambda(\Sigma_{\mu\nu}+\Sigma_{\nu\mu})\;,
\end{eqnarray}
and
\begin{eqnarray}
\lambda&=&\Sigma_{\mu}^{\phantom{\mu}\mu}\;=\;v_{\mu}k^{\mu}\;.
\end{eqnarray}

\begin{table}[h]
\centering
\begin{tabular}{|c | c c c c c c |}
	\hline
 & $\theta^{\alpha}_{\phantom{\alpha}\nu}$ & $\omega^{\alpha}_{\phantom{\alpha}\nu}$ &$S^{\alpha}_{\phantom{\alpha}\nu}$ & $\Lambda^{\alpha}_{\phantom{\alpha}\nu}$ & $\Sigma^{\alpha}_{\phantom{\alpha}\nu}$ & $\Sigma^{\phantom{\nu}\alpha}_{\nu}$ \\
	\hline 
$\theta_{\mu\alpha}$ & $\theta_{\mu\nu}$ & $0$ & $S_{\mu\nu}$ & $\Lambda_{\mu\nu}-\frac{\lambda}{k^2}\Sigma_{\nu\mu}$ & $\Sigma_{\mu\nu}-\lambda\omega_{\mu\nu}$ & $0$ \\ 
$\omega_{\mu\alpha}$ & $0$ & $\omega_{\mu\nu}$ & $0$& $\frac{\lambda}{k^2}\Sigma_{\nu\mu}$ & $\lambda\omega_{\mu\nu}$ & $\Sigma_{\nu\mu}$\\ 
$S_{\mu\alpha}$ & $S_{\mu\nu}$ & $0$ & $-f_{\mu\nu}$& $0$ & $0$ & $0$\\ 
$\Lambda_{\mu\alpha}$ & $\Lambda_{\mu\nu}-\frac{\lambda}{k^2}\Sigma_{\mu\nu}$ & $\frac{\lambda}{k^2}\Sigma_{\mu\nu}$ & $0$& $v^2\Lambda_{\mu\nu}$ & $v^2\Sigma_{\mu\nu}$ & $\lambda\Lambda_{\mu\nu}$\\ 
$\Sigma_{\mu\alpha}$ & $0$ & $\Sigma_{\mu\nu}$ & $0$& $\lambda\Lambda_{\mu\nu}$ & $\lambda\Sigma_{\mu\nu}$ & $k^2\Lambda_{\mu\nu}$\\ 
$\Sigma_{\alpha\mu}$ & $\Sigma_{\nu\mu}-\lambda\omega_{\mu\nu}$ & $\lambda\omega_{\mu\nu}$ & $0$& $v^2\Sigma_{\nu\mu}$ & $v^2k^2\omega_{\mu\nu}$ & $\lambda\Sigma_{\nu\mu}$\\ 
\hline 
\end{tabular}
\caption{Multiplicative table fulfilled by $\theta$, $\omega$, $S$, $\Sigma$ and $\Lambda$. The products obey the
order ``row times column".}
\label{tableA1}
\end{table}

We are now ready to compute the gauge propagator, \textit{i.e.}, the inverse of the wave operator
\begin{eqnarray}
\langle A^a_{\mu}(k)A^b_{\nu}(-k)\rangle&=&i(\mathcal{O}^{-1})^{ab}_{\mu\nu}\;,
\end{eqnarray}
satisfying
\begin{eqnarray} 
\mathcal{O}^{ac}_{\phantom{ac}\mu\alpha}(\mathcal{O}^{-1})_{c\phantom{b\alpha}\nu}^{\phantom{c}b\alpha}&=&\delta^{ab}\left(\theta_{\mu\nu}+\omega_{\mu\nu}\right)\;.
\end{eqnarray}
In fact, for the propagator in the Landau gauge ($\xi=0$), a straightforward computation leads to 
\begin{eqnarray}
\langle A^a_{\mu}(k)A^b_{\nu}(-k)\rangle&=&\frac{i\delta^{ab}}{Q(k)}\left\{-(k^2-\mu^2\Delta v^2)\theta_{\mu\nu}-\frac{\mu^2(v_\alpha k^\alpha)^2[\Omega(k^2-\mu^2\Delta v^2)+k^2]}{P(k)}\omega_{\mu\nu}-\mu S_{\mu\nu}+\right.\nonumber\\
&+&\left.\frac{\mu^2(v_\alpha k^\alpha)[\Omega(k^2-\mu^2\Delta v^2)+k^2]}{P(k)}(\Sigma_{\mu\nu}+\Sigma_{\nu\mu})-\frac{\mu^2k^2[\Omega(k^2-\mu^2\Delta v^2)+k^2]}{P(k)}\Lambda_{\mu\nu}\right\}\;,\nonumber\\
\label{15A1}
\end{eqnarray}
where
\begin{eqnarray}
Q(k)&=&(k^2-\mu^2\Delta v^2)^2+\mu^2[v^2k^2-(v_\alpha k^\alpha)^2]\;,\nonumber\\
P(k)&=&k^2(k^2-\mu^2\Delta v^2)-\Omega\mu^2[v^2k^2-(v_\alpha k^\alpha)^2]\;.
\label{15A2}
\end{eqnarray}
We remark that the propagator \eqref{15A1} is transverse, see Appendix \ref{Tgfp}.

It is worth to mention that the main motivation in the use of the Landau gauge is due to the renormalizability analysis of the model, which is investigated in this gauge \cite{Santos:2014lfa}. In this work, the choice of the Landau gauge is the obvious choice due to the rich set of Ward identities, which simplifies the renormalizability analysis of the model. Moreover, the Landau gauge is a simple gauge to work with in the analysis of the propagator and, hence, a nice choice to start with. Obviously, an analysis in the linear covariant gauges would provide a wider framework to analyze the problem. Nevertheless, the linear covariant gauges analysis is left for future investigation.

\section{Causality and physical spectrum analysis}\label{UC}

Before we analyze the spectrum of the model, we shall perform the general setup of our analysis, \textit{i.e.}, the conditions on the coefficients $\alpha$ and $\beta$ to avoid tachyonic and ghost modes. The nature of the background tensor $v_{\mu}$, namely, spacelike and timelike, will be important for this analysis.

To guarantee that there is no propagation of tachyonic modes we must have $m^2>0$ for each simple pole of the propagator $(k^2=m^2)$, where $m$ is the mass of a particle. An alternative way to see whether tachyonic modes are present in the model is from the concepts of group and front velocities\footnote{Frontal velocity is associated with particle/wave actual propagation while phase velocity is not (at least in general cases). In fact, phase velocity can be greater than 1 with no implication in the breaking of causality, see \cite{Sexl:1976pg}. Nevertheless, the frontal velocity is related to the phase velocity by an infinity momentum limit.} \cite{Sexl:1976pg}. The group velocity is defined as
\begin{eqnarray}\label{v-group}
v_g&=&\frac{d k^0}{d|\textbf{k}|}\;,
\end{eqnarray}
with $k^\mu=(k^0,\textbf{k})$ being the momentum 4-vector of the state. We must have $v_g\leqslant 1$, for causality. On the other hand, the front velocity is
\begin{eqnarray}\label{v-front}
v_f&=&\lim_{|\textbf{k}| \rightarrow \infty}\frac{k^0}{|\textbf{k}|}\;.
\end{eqnarray}
We must also have $v_f\leqslant 1$, for causality.

The existence of physical states in the model will be analyzed here at tree-level by means of the saturation of the free propagator with external currents \cite{Veltman}. The satured propagator is given by
\begin{eqnarray}
\Pi&=&J^{*\mu}_a\textrm{Res}\langle A^a_{\mu}(k)A^b_{\nu}(-k)\rangle J^{\nu}_b\;,
\end{eqnarray}
where $\textrm{Res}\langle A^a_{\mu}(k)A^b_{\nu}(-k)\rangle$ is the residue of the propagator evaluated at each simple pole, and $J^{a\mu}$ is an external current, satisfying the conservation condition, $k_\mu J^{a\mu} =0$. The existence of physical states is ensured whenever the imaginary part of the satured propagator is positive. In summary, we can compute the eigenvalues of the residue matrix at each pole. Whenever the eigenvalues are positive, there is a physical state associated to that pole.

Hence, for causality, we need
\begin{equation}\label{cc}
\bigcap_{i=1}^n m^2_i\;>\;0\;,\;\;\bigcap_{i=1}^n v_{g_i}\;\leqslant\;1\;,\;\;\bigcap_{i=1}^n v_{f_i}\;\leqslant \;1\;,
\end{equation}
where $n$ is the number of poles. For the existence of physical states the eigenvalues of the saturated propagator must satisfy
\begin{equation}\label{uc}
\bigcap_{i=1}^n \lambda_i \geqslant 0\;,
\end{equation}
where the equality stands for massless states.

\subsection{Spacelike case}\label{SLC}

Let us consider first the case where $v^\mu$ is spacelike, \textit{i.e.}, $v^\mu =(0,0,0,1)$ and, without loss of generality, we choose $k^\mu =(k^0,0,0,k^3)$. In the poles $Q(k)=0$ and $P(k)=0$, we have two roots in each pole. Let us start with $Q(k)=0$ by setting $k^2_0=m^2_1$ and $k^2_0=m^2_2$ to obtain the two poles 
\begin{equation}\label{pol1-sp}
m^2_1=k^2_3+(6\alpha+4\beta)\mu^2+\frac{\mu^2}{2}+\frac{\mu}{2}\sqrt{\mu^2+4\left[k^2_3+(6\alpha+4\beta)\mu^2\right]}\;,
\end{equation}
and
\begin{eqnarray}\label{pol2-sp}
m^2_2&=&k^2_3+(6\alpha+4\beta)\mu^2+\frac{\mu^2}{2}-\frac{\mu}{2}\sqrt{\mu^2+4\left[k^2_3+(6\alpha+4\beta)\mu^2\right]}\;.
\end{eqnarray}
The residue matrix in the first pole ($m_1^2$) is
\begin{equation}\label{res1-sp}
R_1=\frac{1}{\mu\sqrt{\mu^2+4\left[k^2_3+(6\alpha+4\beta)\mu^2\right]}}\left(
\begin{array}{ c c c c}
0&0&0&0\\
0&m_1^2-k^2_3-(6\alpha+4\beta)\mu^2&-i\mu m_1&0\\
0&i\mu m_1&m_1^2-k^2_3-(6\alpha+4\beta)\mu^2&0\\
0&0&0&0\\
\end{array}\right)\;.
\end{equation}
This residue matrix presents one nonvanishing eigenvalue
\begin{eqnarray}\label{e1-sp}
\lambda_1&=&1 + \frac{1}{2\sqrt{\frac{1}{4}+\left(\frac{k_3}{\mu}\right)^2+6\alpha+ 4\beta}}\;.
\end{eqnarray}
For the second root of the first pole ($m^2_2$), we  obtain for the residue matrix
\begin{equation}\label{res2-sp}
R_2=-\frac{1}{\mu\sqrt{\mu^2+4\left[k^2_3+(6\alpha+4\beta)\mu^2\right]}}\left(
\begin{array}{ c c c c}
0&0&0&0\\
0&m_2^2-k^2_3-(6\alpha+4\beta)\mu^2&-i\mu m_2&0\\
0&i\mu m_2&m_2^2-k^2_3-(6\alpha+4\beta)\mu^2&0\\
0&0&0&0\\
\end{array}\right)\;,
\end{equation}
also with only one nonvanishing eigenvalue
\begin{eqnarray}\label{e2-sp}
\lambda_2&=&1 - \frac{1}{2\sqrt{\frac{1}{4}+\left(\frac{k_3}{\mu}\right)^2+6\alpha+ 4\beta}}\;.
\end{eqnarray}

Let us now analyze the roots of the pole $P(k)=0$. Setting $k_0^2=m_3^2$ and $k_0^2=m_4^2$ we find for the third and forth poles, respectively,
\begin{eqnarray}\label{pol3-sp}
m^2_3&=&k^2_3+3\alpha\mu^2+\frac{\mu}{2}\sqrt{36\alpha^2\mu^2-16\beta k^2_3}\;,
\end{eqnarray}
and
\begin{eqnarray}\label{pol4-sp}
m^2_4&=&k^2_3+3\alpha\mu^2-\frac{\mu}{2}\sqrt{36\alpha^2\mu^2-16\beta k^2_3}\;.
\end{eqnarray}
The residue matrix of the third pole ($m_3^2$) is
\begin{equation}\label{res3-sp}
R_3=\frac{1}{\mu\sqrt{36\alpha^2\mu^2-16\beta k^2_3}}\left(
\begin{array}{ c c c c}
k^2_3&0&0&-m_3|k_3|\\
0&0&0&0\\
0&0&0&0\\
-m_3|k_3|&0&0&m^2_3\\
\end{array}\right)\;.
\end{equation}
The only nonvanishing eigenvalue is given by
\begin{eqnarray}\label{e3-sp}
\lambda_3&=&\frac{k^2_3+m^2_3}{\mu\sqrt{36\alpha^2\mu^2-16\beta k^2_3}}\;.
\end{eqnarray}
For the fourth pole ($m_4^2$), the corresponding residue matrix is
\begin{equation}\label{res4-sp}
R_4=-\frac{1}{\mu\sqrt{36\alpha^2\mu^2-16\beta k^2_3}}\left(
\begin{array}{ c c c c}
k^2_3 & 0 & 0 &-m_4|k_3| \\
0 & 0 & 0 & 0 \\
0 & 0 & 0 & 0 \\
-m_4|k_3| & 0 & 0 &m^2_4 \\
\end{array}\right)\;.
\end{equation}
This matrix also presents only one nonvanishing eigenvalue
\begin{eqnarray}\label{e4-spa}
\lambda_4&=&-\frac{k^2_3+m^2_4}{\mu\sqrt{36\alpha^2\mu^2-16\beta k^2_3}}\;.
\end{eqnarray}

The group velocities associated to the first two poles ($m^2_1$ and $m^2_2$) are equal,
\begin{eqnarray}\label{e4-spb}
v_{g_1}&=&v_{g_2}\;=\;\frac{1}{\sqrt{1+\frac{1}{4}\left(\frac{\mu}{k_3}\right)^2\left[1+4(6\alpha+4\beta)\right]}}\;,
\end{eqnarray}
and for the remaining two poles ($m^2_3$ and $m^2_4$) are given by
\begin{eqnarray}\label{e4-sp}
v_{g_3}&=&\frac{|k_3|}{\sqrt{k^2_3+3\alpha\mu^2+\mu\sqrt{9\alpha^2\mu^2-4\beta k^2_3}}}\left(1-\frac{2\beta\mu}{\sqrt{9\alpha^2\mu^2-4\beta k^2_3}}\right)\;,\nonumber\\
v_{g_4}&=&\frac{|k_3|}{\sqrt{k^2_3+3\alpha\mu^2-\mu\sqrt{9\alpha^2\mu^2-4\beta k^2_3}}}\left(1+\frac{2\beta\mu}{\sqrt{9\alpha^2\mu^2-4\beta k^2_3}}\right)\;,
\end{eqnarray}
respectively.

Before we study the general case, we analyze two special limit cases by setting $\alpha$ and $\beta$  to zero, not simultaneously.

\subsubsection{The case $\alpha=0$}

The first situation we analyze is the one where we set $\alpha$ to zero. In this case, the poles \eqref{pol1-sp}, \eqref{pol2-sp}, \eqref{pol3-sp}, and \eqref{pol4-sp} reduce to
\begin{eqnarray}\label{p-alpha-0}
m^2_1&=&k^2_3+4\beta\mu^2+\frac{\mu^2}{2}+\mu\sqrt{\frac{\mu^2}{4}+k^2_3+4\beta\mu^2}~,\nonumber\\
m^2_2&=&k^2_3+4\beta\mu^2+\frac{\mu^2}{2}-\mu\sqrt{\frac{\mu^2}{4}+k^2_3+4\beta\mu^2}~,\nonumber\\
m^2_3 &=&k^2_3+2\mu |k_3|\sqrt{-\beta}~,\nonumber\\
m^2_4&=&k^2_3-2\mu |k_3|\sqrt{-\beta}~.
\end{eqnarray}
The new residue matrices \eqref{res1-sp}, \eqref{res2-sp}, \eqref{res3-sp} and \eqref{res4-sp} modify accordingly. The same occurs with the eigenvalues \eqref{e1-sp}, \eqref{e2-sp}, \eqref{e3-sp}, and \eqref{e4-spa} which are listed below,
\begin{eqnarray}\label{e-alpha-0}
\lambda_1&=&1+\frac{1}{2\sqrt{\frac{1}{4}+\left(\frac{k_3}{\mu}\right)^2+4\beta}}\;,\nonumber\\
\lambda_2&=&1-\frac{1}{2\sqrt{\frac{1}{4}+\left(\frac{k_3}{\mu}\right)^2+4\beta}}\;,\nonumber\\
\lambda_3&=&\frac{\left(m_3^2+k_3^2\right)}{4\mu |k_3|\sqrt{-\beta}}\;,\nonumber\\
\lambda_4&=&-\frac{\left(m_4^2+k_3^2\right)}{4\mu |k_3|\sqrt{-\beta}}\;,
\end{eqnarray}
and the group velocities simplify to
\begin{eqnarray}\label{vg-alpha-0}
v_{g_1}&=&v_{g_2}\;=\;\frac{1}{\sqrt{1+\left(\frac{\mu}{k_3}\right)^2\left(\frac{1}{4}+4\beta\right)}}~,\nonumber\\
v_{g_3}&=&\frac{1}{\sqrt{1+2\frac{\mu}{|k_3|}\sqrt{-\beta}}}\left(1+\frac{\mu}{|k_3|}\sqrt{-\beta}\right)~,\nonumber\\
v_{g_4}&=&\frac{1}{\sqrt{1-2\frac{\mu}{|k_3|}\sqrt{-\beta}}}\left(1-\frac{\mu}{|k_3|}\sqrt{-\beta}\right)~.
\end{eqnarray}

\textit{Causality}

In attending criteria \eqref{cc} we have
\begin{eqnarray}\label{clist-m-alpha-0}
\beta\geqslant-\frac{1}{16}\left[1+4\left(\frac{k_3}{\mu}\right)^2\right]~&\Rightarrow & m_1^2>0~,\nonumber\\
\beta\geqslant-\frac{1}{16}\left[1+4\left(\frac{k_3}{\mu}\right)^2\right]~\cap~ \beta\neq-\frac{1}{4}\left(\frac{k_3}{\mu}\right)^2&\Rightarrow & m_2^2>0~,\nonumber\\
\beta\leqslant 0 &\Rightarrow & m_3^2>0~,\nonumber\\
-\frac{1}{4}\left(\frac{k_3}{\mu}\right)^2<\beta\leqslant 0 &\Rightarrow & m_4^2>0~,
\end{eqnarray}
where, for the first two poles, we have employed the triangle inequality. Hence, by intersecting all inequalities in \eqref{clist-m-alpha-0}, we obtain
\begin{equation}\label{cc-alpha-0}
-\frac{1}{4}\left(\frac{k_3}{\mu}\right)^2<\beta \leqslant 0~.
\end{equation}
The condition $v_g\leqslant1$ about group velocities for each mass are satisfied as follows
\begin{eqnarray}\label{vg-list-alpha-0}
\beta\geqslant - \frac{1}{16} &\Rightarrow & v_{g_1}\leqslant 1~\cap~v_{g_2}\leqslant 1~.
\end{eqnarray}
On the other hand, the only way to have $v_{g_3}\leqslant 1$ and $v_{g_4}\leqslant 1$ is $\beta>0$, which implies on imaginary poles. Hence, we have causality violation for the pole $P(k)=0$ at $\alpha=0$.

\textit{Existence of physical states}

It is possible to find restrictions such that all eigenvalues allow the existence of physical states, namely,
\begin{eqnarray}\label{ulist-alpha-0}
\beta \geqslant -\frac{1}{16}\left[1+4\left(\frac{k_3}{\mu}\right)^2\right] &\Rightarrow &\lambda_1 > 0~,\nonumber\\
\beta > -\frac{1}{4}\left(\frac{k_3}{\mu}\right)^2 &\Rightarrow & \lambda_2 > 0~,\nonumber\\
\beta < 0  &\Rightarrow & \lambda_3 > 0~,\nonumber\\
\beta < -\left(\frac{k_3}{\mu}\right)^2  &\Rightarrow & \lambda_4 > 0~.
\end{eqnarray}
However, we do not find an intersection between the inequalities \eqref{ulist-alpha-0}. Hence, ghost modes are always present for $\alpha=0$.

\textit{Causality and physical states}

Accordingly to \eqref{vg-list-alpha-0} and \eqref{ulist-alpha-0}, causality and unitarity are simultaneously violated for $\alpha=0$.

\subsubsection{The case $\beta=0$}

When we set a vanishing $\beta$, the poles turn out to be
\begin{eqnarray}\label{p-beta-0}
m^2_1&=&k^2_3+6\alpha\mu^2+\frac{\mu^2}{2}+\mu^2\sqrt{\frac{1}{4}+\left(\frac{k_3}{\mu}\right)^2+6\alpha}~,\nonumber\\
m^2_2&=&k^2_3+6\alpha\mu^2+\frac{\mu^2}{2}-\mu^2\sqrt{\frac{1}{4}+\left(\frac{k_3}{\mu}\right)^2+6\alpha}~,\nonumber\\
m^2_3&=&k_3^2+6\alpha\mu^2~,\nonumber\\
m^2_4&=&k_3^2~.
\end{eqnarray}
The new eigenvalues are
\begin{eqnarray}\label{e-beta-0-sp}
\lambda_1&=&1+\frac{1}{2\sqrt{\frac{1}{4}+\left(\frac{k_3}{\mu}\right)^2+6\alpha}}~,\nonumber\\
\lambda_2&=&1-\frac{1}{2\sqrt{\frac{1}{4}+\left(\frac{k_3}{\mu}\right)^2+6\alpha}}~,\nonumber\\
\lambda_3&=&1+\frac{1}{3\alpha}\left(\frac{k_3}{\mu}\right)^2~,\nonumber\\
\lambda_4&=&-\frac{1}{3\alpha}\left(\frac{k_3}{\mu}\right)^2~.
\end{eqnarray}
And the new group velocities associated to the $m_1^2$, $m_2^2$, $m_3^2$ and $m_4^2$ are
\begin{eqnarray}
v_{g_1}&=&v_{g_2}\;=\;\frac{1}{\sqrt{1+\frac{1}{4}\left(\frac{\mu}{k_3}\right)^2\left(1+24\alpha\right)}}~,\nonumber\\
v_{g_3}&=&\frac{1}{\sqrt{1+6\alpha\left(\frac{\mu}{k_3}\right)^2}}~,\nonumber\\
v_{g_4}&=&1~,
\end{eqnarray}
respectively.

\textit{Causality}

In order to assure mass positivity we obtain
\begin{eqnarray}\label{clist-m-beta-0}
\alpha\geqslant-\frac{1}{24}\left[1+4\left(\frac{k_3}{\mu}\right)^2\right]~\cap~ \alpha\neq -\frac{1}{6}\left(\frac{k_3}{\mu}\right)^2 &\Rightarrow & m_1^2>0~\cap~ m_2^2>0~,\nonumber\\
\alpha>-\frac{1}{6}\left(\frac{k_3}{\mu}\right)^2 &\Rightarrow & m_3^2>0~,\nonumber\\
\forall\alpha &\Rightarrow & m_4^2>0~.
\end{eqnarray}
As before, we have employed the triangle inequality. To attend the conditions \eqref{cc} we demand
\begin{equation}\label{cc-beta-0}
\alpha>-\frac{1}{6}\left(\frac{k_3}{\mu}\right)^2~.
\end{equation}
The restrictions for the $\alpha$ parameter from group velocities are the following:
\begin{eqnarray}\label{vg-list-beta-0}
\alpha\geqslant -\frac{1}{24} &\Rightarrow & v_{g_1}\leqslant 1~\cap~ v_{g_2}\leqslant 1~,\nonumber\\
\alpha\geqslant 0 &\Rightarrow & v_{g_3}\leqslant 1~,\nonumber\\
\forall\alpha &\Rightarrow & v_{g_4}=1~.
\end{eqnarray}
An intersection between all relations \eqref{vg-list-beta-0} yields in
\begin{equation}\label{vg-beta-0}
\alpha\geqslant 0~.
\end{equation}
Because the solution for $\alpha$ from mass positivity depends on the energy scale, the solution for $\alpha$ from group velocities is more restrictive and is, however, independent from the energy scale.

It remains to infer whether the interval for the front velocities satisfy the causality condition. In fact, conditions \eqref{cc} on the front velocities, for each pole, are satisfied as follows
\begin{equation}\label{vf-list-beta-0}
v_{f_1}\;=\;v_{f_2}\;=\;v_{f_3}\;=\;v_{f_4}\;=\;1~.
\end{equation}
Here, we assume that $|k_3|$ goes to infinity faster than the parameter $\alpha$. This is a reasonable assumption since we can choose some finite $\alpha$ values satisfying \eqref{vg-beta-0}. Therefore, the range $\alpha\geqslant 0$ ensures the causality of the model in the vanishing $\beta$ case.

\textit{Existence of physical states}

In order to satisfy the condition \eqref{uc} we must have
\begin{eqnarray}\label{ulist-beta-0}
\alpha>-\frac{1}{24}\left[1+4\left(\frac{k_3}{\mu}\right)^2\right] &\Rightarrow & \lambda_1>0~,\nonumber\\
\alpha>-\frac{1}{6}\left(\frac{k_3}{\mu}\right)^2 &\Rightarrow & \lambda_2>0~,\nonumber\\
\alpha<-\frac{1}{3}\left(\frac{k_3}{\mu}\right)^2~\cup~ \alpha > 0  &\Rightarrow & \lambda_3>0~,\nonumber\\
\alpha<0 &\Rightarrow & \lambda_4>0~.
\end{eqnarray}
However, the intersection of all regions \eqref{ulist-beta-0} gives no valid range. Hence, the case $\beta=0$ always carries ghost modes.

\textit{Causality and physical states}

We find that, for vanishing $\beta$, causality is ensured for $\alpha\geqslant0$. However, ghost modes always appear. In particular, maintaining $\alpha\geqslant0$, we see from \eqref{ulist-beta-0} that we have three physical massive modes and one massless ghost mode. Hence, there is a chance of full unitarity by considering also the Faddeev-Popov ghosts.

It is worth mentioning that, from \eqref{6}, the mass term proportional to $\alpha$ does not violate Lorentz symmetry, in contrast to the mixing term proportional to $\beta$ which is not invariant under particle Lorentz transformation. In fact, the mass term related to $\alpha$ works like a usual Proca term. On the other hand, the contribution from $\beta$ brings an unusual mixing term, $V^{\mu\nu}A^a_{\mu}A^a_{\nu}$.

\subsubsection{General case}

Now we make the analysis for generic real values of $\alpha$ and $\beta$. Moreover, $|k_3|$ and $\mu$ also have positive values.

\textit{Causality}

We list below the individual conditions for positive masses
\begin{eqnarray}\label{clist-gen}
\beta\geqslant-\frac{1}{16}\left[1+4\left(\frac{k_3}{\mu}\right)^2+24\alpha\right]&\Rightarrow & m_1^2>0~,\nonumber\\
\beta\geqslant-\frac{1}{16}\left[1+4\left(\frac{k_3}{\mu}\right)^2+24\alpha\right]~\cap~\beta\neq -\frac{1}{4}\left[\left(\frac{k_3}{\mu}\right)^2+6\alpha\right] &\Rightarrow & m_2^2>0~,\nonumber\\
\left\{\alpha\leqslant -\frac{1}{3}\left(\frac{k_3}{\mu}\right)^2~\cap~ \beta<-\frac{1}{4}\left[\left(\frac{k_3}{\mu}\right)^2+6\alpha\right]\right\}&~\cup~&\nonumber\\ \left[\alpha>-\frac{1}{3}\left(\frac{k_3}{\mu}\right)^2~\cap~ \beta\leqslant\frac{9\alpha^2}{4\left(\frac{k_3}{\mu}\right)^2}\right] &\Rightarrow & m_3^2>0~,\nonumber\\
\alpha>-\frac{1}{3}\left(\frac{k_3}{\mu}\right)^2~\cap~ -\frac{1}{4}\left[\left(\frac{k_3}{\mu}\right)^2+6\alpha\right]<\beta\leqslant\frac{9\alpha^2}{4\left(\frac{k_3}{\mu}\right)^2} &\Rightarrow & m_4^2>0~.\nonumber\\
\end{eqnarray}
Hence, the intersection of the above inequalities is
\begin{equation}\label{cc-k3-mub}
\alpha>-\frac{1}{3}\left(\frac{k_3}{\mu}\right)^2~\cap~-\frac{1}{4}\left[\left(\frac{k_3}{\mu}\right)^2+6\alpha\right]<\beta\leqslant \frac{9\alpha^2}{4\left(\frac{k_3}{\mu}\right)^2}~,
\end{equation}
which is the unique interval that obeys the condition \eqref{cc}.

The intervals for the group velocities are quite complicated. Indeed, the general solution is possible to be found, see Appendix \ref{SCS}. Nevertheless, a simple analytical solution can be obtained,
\begin{eqnarray}\label{vg-list-gen}
\beta\geqslant -\frac{1}{16}\left(1+24\alpha\right) \;\;\Rightarrow \;\; v_{g_1}\leqslant 1 &\cap & v_{g_2}\leqslant 1,\nonumber\\
\beta=-\frac{3}{2}\alpha\;\;~\cap~\;\;\alpha>0\;\;\Rightarrow \;\; v_{g_3}\leqslant 1 &\cap& v_{g_4}\leqslant 1.
\end{eqnarray}
The intersection of \eqref{vg-list-gen} reads
\begin{eqnarray}\label{vg-list-gen1}
\beta=-\frac{3}{2}\alpha\;\;~\cap~\;\;\alpha>0\;.
\end{eqnarray}
The front velocities for each mass are easily obtained,
\begin{eqnarray}\label{vf-list-gen}
v_{f_1}\;=\;v_{f_2}\;=\;v_{f_3}\;=\;v_{f_4}\;=\;1~,
\end{eqnarray}
and they satisfy the conditions listed in \eqref{cc}. Again, we are assuming that the $\alpha$ parameter is kept finite as $|k_3|$ increases.

\textit{Existence of physical states}

Turning back to the general case, ghosts modes will be avoided for
\begin{eqnarray}\label{ulist-gen}
\beta>-\frac{1}{16}\left[1+4\left(\frac{k_3}{\mu}\right)^2+24\alpha\right] &\Rightarrow & \lambda_1>0~,\nonumber\\
\beta> -\frac{1}{4}\left[\left(\frac{k_3}{\mu}\right)^2+6\alpha\right] &\Rightarrow & \lambda_2>0~,\nonumber\\
\left\{\alpha\leqslant -\frac{2}{3}\left(\frac{k_3}{\mu}\right)^2~\cap~ \beta<-\left[\left(\frac{k_3}{\mu}\right)^2+3\alpha\right]\right\}\cup\left[\alpha> -\frac{2}{3}\left(\frac{k_3}{\mu}\right)^2~\cap~\beta<\frac{9\alpha^2}{4\left(\frac{k_3}{\mu}\right)^2}\right]&\Rightarrow & \lambda_3>0~,\nonumber\\
\left[\alpha\leqslant -\frac{2}{3}\left(\frac{k_3}{\mu}\right)^2~\cap~ \beta<\frac{9\alpha^2}{4\left(\frac{k_3}{\mu}\right)^2}\right]\cup\left\{\alpha> -\frac{2}{3}\left(\frac{k_3}{\mu}\right)^2~\cap~ \beta<-\left[\left(\frac{k_3}{\mu}\right)^2+3\alpha\right]\right\} &\Rightarrow & \lambda_4>0~.\nonumber\\
\end{eqnarray}
The intersection of all inequalities in \eqref{ulist-gen} is
\begin{equation}\label{ulist-genb}
\alpha<-\frac{1}{2}\left(\frac{k_3}{\mu}\right)^2~\cap~-\frac{1}{4}\left[\left(\frac{k_3}{\mu}\right)^2+6\alpha\right]<\beta<-\left[\left(\frac{k_3}{\mu}\right)^2+3\alpha\right]~.
\end{equation}

\textit{Causality and physical states}

There is no valid range in the intersection of \eqref{ulist-genb} and those in Appendix \ref{SCS}. It can also be checked for the particular case \eqref{vg-list-gen1}. Nevertheless, \eqref{vg-list-gen1} and \eqref{ulist-genb} provide three massive physical modes and one massive ghost mode.

\subsection{Timelike case}\label{TLC}

In this section we study the causality and the existence of physical states of the model for a timelike background vector, \textit{i.e.}, $v^\mu =(1,0,0,0)$. We follow the same strategy as in the spacelike case. The four roots for the two poles $Q(k)=0$ and $P(k)=0$ are given by
\begin{eqnarray}
m^2_1&=&k^2_3-(6\alpha+4\beta)\mu^2+\mu|k_3|\;,\nonumber\\
m^2_2&=&k^2_3-(6\alpha+4\beta)\mu^2-\mu|k_3|\;,\nonumber\\
m^2_3&=&k^2_3-(3\alpha+2\beta)\mu^2+\mu\sqrt{\mu^2(3\alpha+2\beta)^2-4\beta k^2_3}\;,\nonumber\\
m^2_4&=&k^2_3-(3\alpha+2\beta)\mu^2-\mu\sqrt{\mu^2(3\alpha+2\beta)^2-4\beta k^2_3}\;,
\label{AA1}
\end{eqnarray}
where the roots $m^2_1$ and $m^2_2$ correspond to the pole $Q(k)=0$, and the roots $m^2_3$ and $m^2_4$ are related to the pole $P(k)=0$. Thus, we can evaluate the residue matrices of the propagator \eqref{15A1} when $k^2_0$ assumes the values $m^2_1$, $m^2_2$, $m^2_3$, and $m^2_4$. The result is quite simple for $m^2_1$ and $m^2_2$:
\begin{equation}
R_1=\frac{1}{2}\left(
\begin{array}{ c c c c}
0&0&0&0\\
0&1&-i&0\\
0&i&1&0\\
0&0&0&0\\
\end{array}\right)\;,\;\;\;\;\lambda_1=1\;.
\label{AA2}
\end{equation}
\begin{equation}
R_2=\frac{1}{2}\left(
\begin{array}{ c c c c}
0&0&0&0\\
0&1&i&0\\
0&-i&1&0\\
0&0&0&0\\
\end{array}\right)\;,\;\;\;\;\;\lambda_2=1\;.
\label{AA3}
\end{equation} 
For $m_3^2$ we find
\begin{equation}
R_3=\frac{1}{\mu\sqrt{\mu^2(6\alpha+4\beta)^2-16\beta k^2_3}}\left(
\begin{array}{ c c c c}
k^2_3&0&0&-m_3|k_3|\\
0&0&0&0\\
0&0&0&0\\
-m_3|k_3|&0&0&m^2_3\\
\end{array}\right)\ ,
\label{AA4}
\end{equation}
with a single nonvanishing eigenvalue given by
\begin{eqnarray}
\lambda_3&=&\frac{k^2_3+m^2_3}{\mu\sqrt{\mu^2(6\alpha+4\beta)^2-16\beta k^2_3}}\;.
\end{eqnarray}
For $m_4^2$ we find
\begin{equation}
R_4=-\frac{1}{\mu\sqrt{\mu^2(6\alpha+4\beta)^2-16\beta k^2_3}}\left(
\begin{array}{ c c c c}
k^2_3&0&0&-m_4|k_3|\\
0&0&0&0\\
0&0&0&0\\
-m_4|k_3|&0&0&m^2_4\\
\end{array}\right)\;,
\label{AA6}
\end{equation}
with a single nonvanishing eigenvalue given by
\begin{eqnarray}
\lambda_4&=&-\frac{k^2_3+m^2_4}{\mu\sqrt{\mu^2(6\alpha+4\beta)^2-16\beta k^2_3}}\;.
\end{eqnarray}

The group velocities associated to each pole are given by
\begin{eqnarray}
v_{g_1}&=&\left(|k_3|+\frac{\mu}{2}\right)\frac{1}{\sqrt{k^2_3-(6\alpha+4\beta)\mu^2+\mu|k_3|}}\;,\nonumber\\
v_{g_2}&=&\left(|k_3|-\frac{\mu}{2}\right)\frac{1}{\sqrt{k^2_3-(6\alpha+4\beta)\mu^2-\mu|k_3|}}\;,\nonumber\\
v_{g_3}&=&\left(1-\frac{2\beta\mu}{\sqrt{\mu^2(3\alpha+2\beta)^2-4\beta k^2_3}}\right)\frac{|k_3|}{\sqrt{k^2_3-(3\alpha+2\beta)\mu^2+\mu\sqrt{\mu^2(3\alpha+2\beta)^2-4\beta k^2_3}}}\;,\nonumber\\
v_{g_4}&=&\left(1+\frac{2\beta\mu}{\sqrt{\mu^2(3\alpha+2\beta)^2-4\beta k^2_3}}\right)\frac{|k_3|}{\sqrt{k^2_3-(3\alpha+2\beta)\mu^2-\mu\sqrt{\mu^2(3\alpha+2\beta)^2-4\beta k^2_3}}}\;.\nonumber\\
\end{eqnarray}

\textit{Causality}

The positive-definiteness for the poles are found to be
\begin{eqnarray}\label{ecTL}
\beta<\frac{1}{4}\left[\left(\frac{k_3}{\mu}\right)^2+\frac{|k_3|}{\mu}-6\alpha\right]&\Rightarrow & m_1^2>0~,\nonumber\\
\beta<\frac{1}{4}\left[\left(\frac{k_3}{\mu}\right)^2-\frac{|k_3|}{\mu}-6\alpha\right]&\Rightarrow & m_2^2>0~,\nonumber\\
\left\{
\alpha<\frac{1}{6}\left(\frac{k_3}{\mu}\right)^2 ~\cap~\beta\leqslant\frac{1}{2}\left[\left(\frac{k_3}{\mu}\right)^2-3\alpha\right]-\frac{1}{2}\frac{|k_3|}{\mu}\sqrt{\left(\frac{k_3}{\mu}\right)^2-6\alpha}\right\}~\cup\nonumber\\
\left\{\alpha=\frac{1}{6}\left(\frac{k_3}{\mu}\right)^2~\cap~ \beta<\frac{1}{2}\left[\left(\frac{k_3}{\mu}\right)^2-3\alpha\right]-\frac{1}{2}\frac{|k_3|}{\mu}\sqrt{\left(\frac{k_3}{\mu}\right)^2-6\alpha}\right\}~\cup\nonumber\\
\left\{\alpha> \frac{1}{6}\left(\frac{k_3}{\mu}\right)^2~\cap~\forall\beta\right\}  &\Rightarrow & m_3^2>0~,\nonumber\\
\alpha<\frac{1}{6}\left(\frac{k_3}{\mu}\right)^2 ~\cap~\beta\leqslant\frac{1}{2}\left[\left(\frac{k_3}{\mu}\right)^2-3\alpha\right]-\frac{1}{2}\frac{|k_3|}{\mu}\sqrt{\left(\frac{k_3}{\mu}\right)^2-6\alpha}&\Rightarrow & m_4^2>0~.\nonumber\\
\end{eqnarray}
Inequalities \eqref{ecTL}, for $|k_3|\ne\mu$, provides
\begin{eqnarray}\label{ec-intersec-TL}
& &\left\{\alpha<-\frac{1}{24}\left[1+2\frac{|k_3|}{\mu}-3\left(\frac{k_3}{\mu}\right)^2\right]~\cap  ~ \beta\leqslant\frac{1}{2}\left[\left(\frac{k_3}{\mu}\right)^2-3\alpha\right]-\frac{1}{2}\frac{|k_3|}{\mu}\sqrt{\left(\frac{k_3}{\mu}\right)^2-6\alpha}\right\}~\cup\nonumber\\
&&\cup~\left\{\alpha=-\frac{1}{24}\left[1+2\frac{|k_3|}{\mu}-3\left(\frac{k_3}{\mu}\right)^2\right]~\cap ~\beta<\frac{1}{2}\left[\left(\frac{k_3}{\mu}\right)^2-3\alpha\right]-\frac{1}{2}\frac{|k_3|}{\mu}\sqrt{\left(\frac{k_3}{\mu}\right)^2-6\alpha}\right\}~\cup\nonumber\\
&&\cup ~\left\{-\frac{1}{24}\left[1+2\frac{|k_3|}{\mu}-3\left(\frac{k_3}{\mu}\right)^2\right]<\alpha<\frac{1}{6}\left(\frac{k_3}{\mu}\right)^2 ~\cap~ \beta<\frac{1}{4}\left[-\frac{|k_3|}{\mu}+\left(\frac{k_3}{\mu}\right)^2-6\alpha\right]\right\}~,\nonumber\\
\end{eqnarray}
while for $|k_3|/\mu=1$, we obtain
\begin{eqnarray}\label{et}
\left\{\alpha<0~\cap  ~ \beta\leqslant\frac{1}{2}\left(1-3\alpha\right)-\frac{1}{2}\sqrt{1-6\alpha}\right\}\cup ~\left\{0\leqslant\alpha<\frac{1}{6}~\cap~\beta<-\frac{3\alpha}{2}\right\}~.
\label{ec}
\end{eqnarray}

The causal intervals for the group velocities are
\begin{eqnarray}\label{vg-bvben}
\beta\leqslant -\frac{1}{16}\left(1+24\alpha\right)\;\; \Rightarrow \;\; v_{g_1}\leqslant 1&~\cap~&v_{g_2}\leqslant 1\;,\nonumber\\
\left\{\alpha\leqslant 0~\cap~\beta\leqslant 0\right\}&\cup&\nonumber\\
\left\{0<\alpha <\frac{1}{6}\left(\frac{k_3}{\mu}\right)^2~\cap~\beta\leqslant\frac{1}{\mathcal{Y}}\left\{\mathcal{W}\left[1+\frac{\alpha}{\mathcal{Z}}\left(\frac{k_3}{\mu}\right)^2\right]+\mathcal{Z}\right\}\right\}~\Rightarrow~v_{g_3}\leqslant 1&~\cap~&v_{g_4}\leqslant 1\;,\nonumber\\
\end{eqnarray}
where
\begin{eqnarray}\label{gveloc-tlk}
\mathcal{Y}\equiv\mathcal{Y}(k_3,\mu,\alpha)&=&2\left[\left(\frac{k_3}{\mu}\right)^2-6\alpha\right]~,\nonumber\\
\mathcal{W}\equiv\mathcal{W}(k_3,\mu,\alpha)&=&2\alpha\left[2\left(\frac{k_3}{\mu}\right)^2-9\alpha\right]~,\nonumber\\
\mathcal{Z}\equiv\mathcal{Z}(k_3,\mu,\alpha)&=&\left\{-8\alpha^3 \left(\frac{k_3}{\mu}\right)^6+135\alpha\left(\frac{k_3}{\mu}\right)^4-486\alpha^5\left(\frac{k_3}{\mu}\right)^2+\right.\nonumber\\
&+&\left. 9\left(\frac{k_3}{\mu}\right)^2\sqrt{\alpha^7\left[81\alpha-16\left(\frac{k_3}{\mu}\right)^2\right]\left[\left(\frac{k_3}{\mu}\right)^2-6\alpha\right]^2}\right\}^{1/3}~.\nonumber\\
\end{eqnarray}
The intersections of all inequalities in \eqref{gveloc-tlk} provides
\begin{eqnarray}\label{gveloc-tlk-all}
\left\{\alpha\leqslant-\frac{1}{24}~\cap~\beta\leqslant 0\right\}~&\cup&~\nonumber\\
\left\{-\frac{1}{24}<\alpha\leqslant 0~\cap~\beta\leqslant-\frac{1}{16}\left(1+24\alpha\right)\right\} ~&\cup&~\nonumber\\
\left\{0<\alpha<\frac{1}{24}~\cap~\left[\beta\leqslant-\frac{1}{16}\left(1+24\alpha\right)\cap\frac{|k_3|}{\mu}\geqslant\sqrt{\frac{3\alpha}{4\beta}\left(3\alpha+2\beta\right)}\right]\right\} ~&\cup&~\nonumber\\
\left\{\alpha\geqslant\frac{1}{24}~\cap~\beta<-3\alpha~\cap~\frac{|k_3|}{\mu}\geqslant\sqrt{\frac{3\alpha}{4\beta}\left(3\alpha+2\beta\right)}\right\}~.
\end{eqnarray}

And, finally, for the front velocities we find
\begin{equation}\label{vfbhjh-gen}
v_{f_1}\;=\;v_{f_2}\;=\;v_{f_3}\;=\;v_{f_4}\;=\;1~,
\end{equation}
and they satisfy the conditions listed in \eqref{cc}.

\textit{Existence of physical states}

The poles $m_1^2$ and $m_2^2$ are automatically physical states. To avoid ghost modes in the third pole, we need
\begin{eqnarray}\label{utlk-pole3}
\left\{\alpha\leq 0~\cap~\left\{\beta<\frac{1}{2}\left[\left(\frac{k_3}{\mu}\right)^2-3\alpha-\sqrt{\left(\frac{k_3}{\mu}\right)^4-6\alpha \left(\frac{k_3}{\mu}\right)^2}\right]~\cup~\beta>\left(\frac{k_3}{\mu}\right)^2-3\alpha\right\}\right\}~&\cup&~\nonumber\\
\left\{0<\alpha<\frac{1}{6}\left(\frac{k_3}{\mu}\right)^2~\cap~\left\{\beta<\frac{1}{2}\left[\left(\frac{k_3}{\mu}\right)^2-3\alpha-\sqrt{\left(\frac{k_3}{\mu}\right)^4-6\left(\frac{k_3}{\mu}\right)^2\alpha}\right]\right.\right.~&\cup&~\nonumber\\
\left.\left.\beta>\frac{1}{2}\left[\left(\frac{k_3}{\mu}\right)^2-3\alpha+\sqrt{\left(\frac{k_3}{\mu}\right)^4-6\left(\frac{k_3}{\mu}\right)^2\alpha}\right]\right\}\right\}~&\cup&~\nonumber\\
\left\{\alpha=\frac{1}{6}\left(\frac{k_3}{\mu}\right)^2~\cap~\beta<\frac{1}{2}\left[\left(\frac{k_3}{\mu}\right)^2-3\alpha-\sqrt{\left(\frac{k_3}{\mu}\right)^4-6\left(\frac{k_3}{\mu}\right)^2\alpha}\right]\right.~&\cup&\nonumber\\
\left.\beta>\frac{1}{2}\left[\left(\frac{k_3}{\mu}\right)^2-3\alpha-\sqrt{\left(\frac{k_3}{\mu}\right)^4-6\left(\frac{k_3}{\mu}\right)^2\alpha}\right]\right\}~&\cup&~\nonumber\\
\left\{\alpha>\frac{1}{6}\left(\frac{k_3}{\mu}\right)^2~\cap~\forall\beta\right\} \Rightarrow \lambda_3>0~,\nonumber\\
\end{eqnarray}
and, for the fourth pole we must have
\begin{eqnarray}\label{utlk-pole4}
\left\{\alpha\leqslant 0~\cap~\beta<\frac{1}{2}\left[\left(\frac{k_3}{\mu}\right)^2-3\alpha+\sqrt{\left(\frac{k_3}{\mu}\right)^4-6\alpha \left(\frac{k_3}{\mu}\right)^2}\right]\right\}&\cup&\nonumber\\
\left\{\alpha>0~\cap~\beta>\left(\frac{k_3}{\mu}\right)^2-3\alpha\right\}&\Rightarrow &\lambda_4>0~.
\end{eqnarray}
The intersection between \eqref{utlk-pole3} and \eqref{utlk-pole4} is simply given by
\begin{equation}\label{utlk-pole-all}
\beta>\left(\frac{k_3}{\mu}\right)^2-3\alpha~.
\end{equation}

\textit{Causality and physical states}

Since there is no solution when we intersect the conditions listed in \eqref{gveloc-tlk-all} and \eqref{utlk-pole-all} then we do not get causality and all four poles as physical states simultaneously. Nevertheless, since we always have three physical states for $m_1^2,\;m^2_2$ and $m_3^2$, we can impose causality by means of \eqref{gveloc-tlk-all} and deal with one massive ghost state.

It is worth to point out that, although these two cases -- spacelike and timelike -- have similar results, the timelike case is a bit more cumbersome due to the point $|k_3|=1$ which behaves differently from any other value.

\subsection{Lightlike case}

In the case of a lightlike background, all terms depending only on the Lorentz violating scale $\mu$ and the dimensionless parameters vanish, since $v^{\mu}=(1,0,0,1)$. Thus, the remaining  terms on the poles $Q(k)=0$ and $P(k)=0$ are respectively $k^2=\pm \mu v_{\alpha}k^{\alpha}$ and $k^2=\pm 2\sqrt{-\beta}\mu v_{\alpha}k^{\alpha}$. Hence, $\beta<0$ must be required. In all four cases, it is easy to check that the group and front velocities are both equal to 1, ensuring causality. In the analysis of the existence of the physical states it is possible to see that the model is plagued with inconsistent modes. In fact, from the roots shown above, we see that all four roots admit a common solution, \textit{i.e.}, $k_0=|k_3|$. In fact, this solution of the poles, combined with the null vector $v^\mu$, provides $v_\alpha k^\alpha=0$ and also $k^2=0$. Hence, we have four simple poles coinciding at the same point in the $k$-space. And this is the essence of the inconsistency.

\section{Conclusions}\label{FINAL}

In this work we have studied the minimal requirements for causality and unitarity for a pure Yang-Mills theory with Lorentz violation by means of the study of the gauge field propagator poles. The propagator of the gauge field is modified due to the mass generation originated from Symanzik sources, BRST quantization and Lorentz violating terms. The mass parameters are related to the dimensionless parameters $\alpha$ and $\beta$ and to the violation parameter $\mu$. Remarkably, the tree-level propagator is transverse (Appendix \ref{Tgfp}). Essentially, we have found the restrictions on these parameters in order to guarantee minimal conditions for physical consistency of the model.

The main results here are: 
\begin{itemize}
\item It was shown that the tree-level propagator is transverse, in despite of the presence of mass parameters.

\item We analyzed the causality and the spectrum of the gauge field in the case of spacelike background field. The situation where causality holds are consistent with three massive physical modes and one massive ghost mode. If $\beta=0$, the ghost mode turns to a massless one.

\item In the timelike case, the causality is consistent with three massive physical states and one massive ghost states.

\end{itemize}

It is worth mentioning that, although the model resembles a Higgs mechanism for the extended QED \cite{BaetaScarpelli:2003yd}, the situation is slightly different. In fact, the mass term generated for the photon field in \cite{BaetaScarpelli:2003yd} comes from Higgs mechanism, without any relation with the Lorentz violation scale $\mu$, and the Goldstone modes are absorbed by the photon in such a way that causality and unitarity of the model are preserved. The same happens for the non-Abelian theory with symmetry group $SO(3)$ with the Higgs field \cite{Baeta Scarpelli:2006kd}. Inhere, the mass term is originated from renormalizability requirements and depends directly on the Lorentz violation scale $\mu$. Furthermore, we have analyzed a non-Abelian model, where the Faddeev-Popov ghosts develop a fundamental role in eliminating the non-physical degrees of freedom. Thus, one-loop explicit computations may be required to see if these modes decouple from the model.

Finally, we remark that there are also possibilities to consider imaginary poles of the propagators. Following the interpretation of the Gribov-Zwanziger approach to QCD \cite{Zwanziger:1989mf,Dudal:2005na,Dudal:2011gd,Capri:2016aqq}, this situation is deeply related to confinement. Hence, it would be interesting to see if the present model is already free of Gribov copies or not due to the presence of Lorentz violation.

\appendix

\section{Some complete results}\label{SCS}

In the general case for the spacelike background, the third group velocity must satisfy the causality condition when $\beta>0$ and
\begin{eqnarray}\label{vgroup3-interv1}
\alpha &\leqslant & -\frac{1}{18}\left[3\beta +\sqrt{3}\left(\sqrt{\mathcal{A}+\mathcal{C}}+\sqrt{2\mathcal{A}-\mathcal{C}+\mathcal{D}}\right)\right] ~\cup\nonumber\\
\alpha &\geqslant & -\frac{1}{18}\left[3\beta +\sqrt{3}\left(\sqrt{\mathcal{A}+\mathcal{C}}-\sqrt{2\mathcal{A}-\mathcal{C}+\mathcal{D}}\right)\right]~.
\end{eqnarray}
Or, if $\beta<0$ then
\begin{equation}\label{vgroup3-interv2}
\alpha \geqslant -\frac{1}{18}\left[3\beta +\sqrt{3}\left(\sqrt{\mathcal{A}+\mathcal{C}}+\sqrt{2\mathcal{A}-\mathcal{C}+\mathcal{D}}\right)\right]~,
\end{equation}
where
\begin{eqnarray}
\mathcal{A}\equiv\mathcal{A}(\beta,k_3,\mu)&=& 8\left(\frac{k_3}{\mu}\right)^2+3\beta^2 ~,\nonumber\\
\mathcal{B}\equiv\mathcal{B}(\beta,k_3,\mu)&=&-\left(\frac{k_3}{\mu}\right)^2\beta^3\left[8\left(\frac{k_3}{\mu}\right)^4+27\beta^2-3\sqrt{3}\beta\sqrt{16\left(\frac{k_3}{\mu}\right)^4+27\beta^2}\right]~,\nonumber\\
\mathcal{C}\equiv\mathcal{C}(\beta,k_3,\mu)&=&2\left[\frac{4\beta^2}{\mathcal{B}(\beta,k_3,\mu)}\left(\frac{k_3}{\mu}\right)^2+\mathcal{B}(\beta,k_3,\mu)\right]~,\nonumber\\
\mathcal{D}\equiv\mathcal{D}(\beta,k_3,\mu)&=&\frac{6\sqrt{3}\left[4\left(\frac{k_3}{\mu}\right)^2-\beta\right]\beta^2}{\sqrt{\mathcal{A}(\beta,k_3,\mu)+\mathcal{B}(\beta,k_3,\mu)}}~.
\end{eqnarray}
Due to this complicated interval, we opted to use the condition $\beta=-3\alpha/2$ with $\alpha>0$ in the text.

For the fourth group velocity the parameters must simply satisfy
\begin{equation}\label{vgroup4-interv}
\beta<0~\cap~\alpha \geqslant -\frac{1}{18}\left[3\beta +\sqrt{3}\left(\sqrt{\mathcal{A}+\mathcal{C}}-\sqrt{2\mathcal{A}-\mathcal{C}+\mathcal{D}}\right)\right]~.
\end{equation}
A more complete interval for the four group velocities satisfying the causality condition \eqref{cc} is the intersection of inequalities \eqref{vgroup3-interv1}, \eqref{vgroup3-interv2} and \eqref{vgroup4-interv}, $i.e.$,
\begin{equation}\label{vgroup-gencase}
-\frac{1}{16}\left( 1+24\alpha\right)\leqslant\beta< 0~\cap ~\alpha \geqslant -\frac{1}{18}\left[3\beta +\sqrt{3}\left(\sqrt{\mathcal{A}+\mathcal{C}}-\sqrt{2\mathcal{A}-\mathcal{C}+\mathcal{D}}\right)\right]~.
\end{equation}

\section{Transversality of the gauge field propagator}\label{Tgfp}

It was shown in \cite{Santos:2014lfa} that the gauge field propagator in the Landau gauge remains transverse to all orders in perturbation theory. Although the gauge propagator considered in \cite{Santos:2014lfa} is the usual gauge propagator for the gluon field, it is expected that the Lorentz violating terms which were controlled by Symanzik sources do not spoil this property, once the gauge fixing Ward identity is not modified by this approach. In fact, when the Symanzik sources attain their physical values\footnote{See \cite{Santos:2014lfa} for extra details.}, the transversality of the modified gauge field propagator still holds. This can be easily seen by rewriting the gauge field propagator \eqref{15A1} as
\begin{eqnarray}
\langle A^a_{\mu}(k)A^b_{\nu}(-k)\rangle&=&\frac{i\delta^{ab}}{Q(k)}\left\{-(k^2-\mu^2\Delta v^2)\theta_{\mu\nu}-\frac{\mu^2k^2[\Omega(k^2-\mu^2\Delta v^2)+k^2]}{P(k)}\theta_{\mu\alpha}\Lambda^{\alpha\beta}\theta_{\beta\nu}-\mu S_{\mu\nu}\right\}\;.\nonumber\\
\end{eqnarray}
Hence, it is straightforward to show that
\begin{eqnarray}
k^{\mu}\langle A^a_{\mu}(k)A^b_{\nu}(-k)\rangle&=&k^{\nu}\langle A^a_{\mu}(k)A^b_{\nu}(-k)\rangle\;=\;0\;.
\end{eqnarray}
This conclusion also applies to the Abelian Lorentz violating theory. See \cite{Kostelecky:2001jc} for the one-loop case and \cite{Santos:2015koa,Santos:2016bqc} for the algebraic proof, which holds at any order in loop expansion.

\section*{Acknowledgements}

The Conselho Nacional de Desenvolvimento Cient\'{i}fico e Tecnol\'{o}gico, The Coordena\c c\~ao de Aperfei\c coamento de Pessoal de N\'ivel Superior (CAPES) and the Pr\'o-Reitoria de Pesquisa, P\'os-Gradua\c c\~ao e Inova\c c\~ao (PROPPI-UFF) are acknowledge for financial support.

\end{document}